\def\rxivcitationstyle{author-date}

\documentclass[times, twoside, watermark]{rxiv_maker_style}
\usepackage{blindtext} 
\usepackage{float} 

\renewcommand{\today}{2025-07-02}

\leadauthor{Saraiva}

\begin{document}
\title{EZInput: A Cross-Environment Python Library for Easy UI Generation in Scientific Computing}

\shorttitle{EZInput}

\author[1,\Letter]{Bruno M. Saraiva}
\author[2]{Iván Hidalgo-Cenalmor}
\author[1]{António D. Brito}
\author[1]{Damián Martínez}
\author[1]{Tayla Shakespeare}
\author[2,3,4,5,\Letter]{Guillaume Jacquemet}
\author[1,6,\Letter]{Ricardo Henriques}
\affil[1]{Instituto de Tecnologia Química e Biológica António Xavier, Universidade Nova de Lisboa, Oeiras, Portugal}
\affil[2]{Faculty of Science and Engineering, Cell Biology, Åbo Akademi University, Turku, Finland}
\affil[3]{InFLAMES Research Flagship Center, University of Turku, Turku, Finland}
\affil[4]{Turku Bioscience Centre, University of Turku and Åbo Akademi University, Turku, Finland}
\affil[5]{Foundation for the Finnish Cancer Institute, Helsinki, Finland}
\affil[6]{UCL Laboratory for Molecular Cell Biology, University College London, London, United Kingdom}

\maketitle

\begin{abstract}

Researchers face a persistent barrier when applying computational algorithms with parameter configuration typically demanding programming skills, interfaces differing across environments, and settings rarely persisting between sessions. This fragmentation forces repetitive input, slows iterative exploration, and undermines reproducibility because parameter choices are difficult to record, share, and reuse.
We present EZInput, a cross-runtime environment 
 Python library enabling algorithm developers to automatically generate graphical user interfaces that make their computational tools accessible to end-users without programming expertise. EZInput employs a declarative specification system where developers define input requirements and validation constraints once; the library then handles environment detection, interface rendering, parameter validation, and session persistence across Jupyter notebooks, Google Colab, and terminal environments. This "write once, run anywhere" architecture enables researchers to prototype in notebooks and deploy identical parameter configurations for batch execution on remote systems without code changes or manual transcription. 
 Parameter persistence, inspired by ImageJ/FIJI and adapted to Python workflows, saves and restores user configurations via lightweight YAML files, eliminating redundant input and producing shareable records that enhance reproducibility. EZInput supports the input types common in scientific computing, with built-in validation and clear feedback.

\end{abstract}


\begin{corrauthor}
(B. M. Saraiva) bsaraiva\at itqb.unl.pt; 
(G. Jacquemet) guillaume.jacquemet\at abo.fi; 
(R. Henriques) r.henriques\at itqb.unl.pt
\end{corrauthor}

\section*{Main}

\begin{figure}[t]
\centering
\makebox[\linewidth][c]{
  \includegraphics[width=\linewidth,keepaspectratio,draft=false]{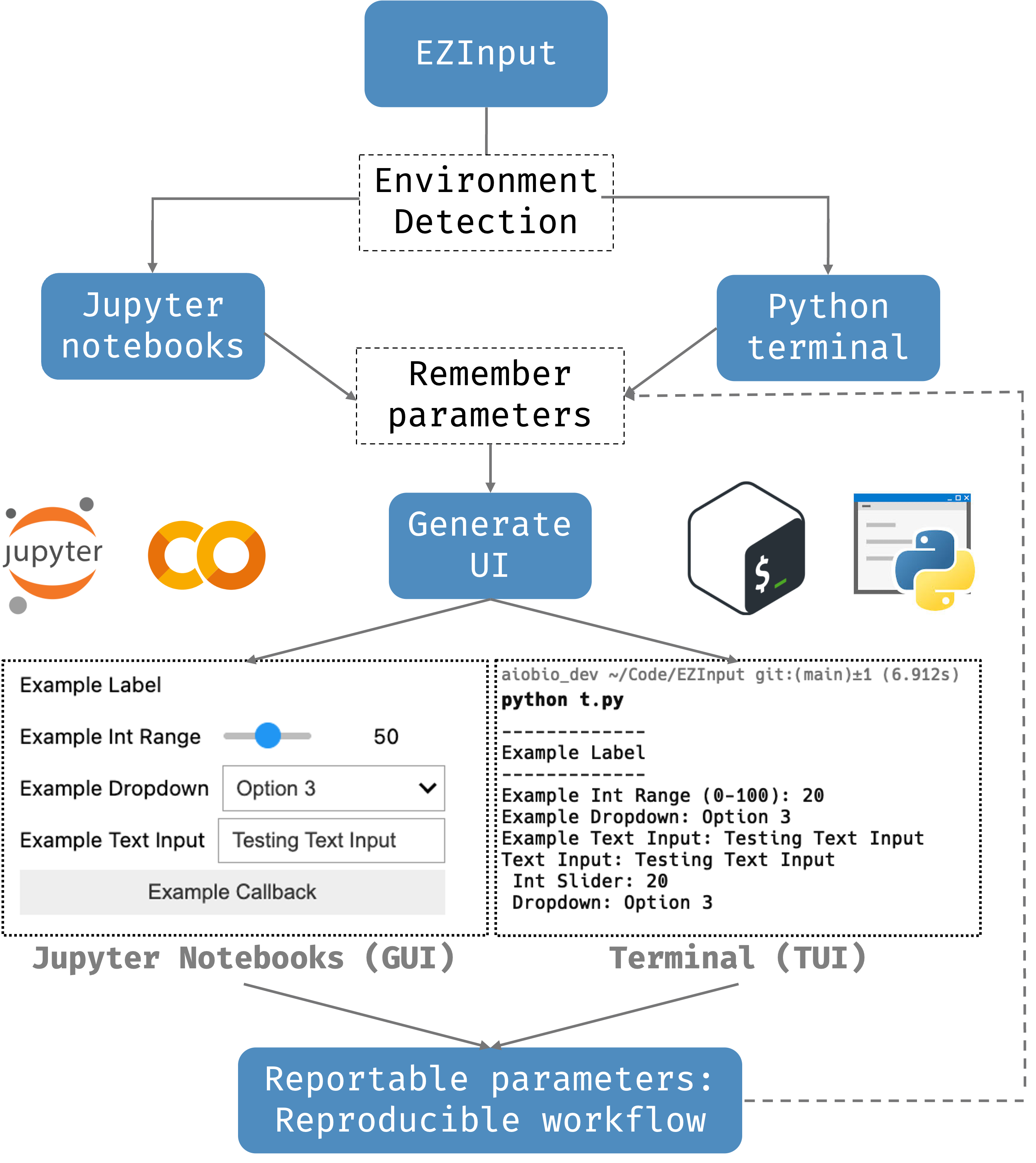}
}
\begingroup\captionsetup{width=\linewidth,singlelinecheck=false,justification=justified}\setlength{\abovecaptionskip}{6pt}\setlength{\belowcaptionskip}{6pt}\ifdefined\justifying\justifying\fi\caption{\textbf{EZInput framework architecture and workflow integration.} The EZInput library implements a declarative parameter specification system that automatically generates graphical user interfaces across multiple computational environments, both Jupyter notebooks and terminal environments, without additional interface development. Parameter persistence mechanism inspired by ImageJ/FIJI \citep{imagej,schindelin2012fiji}, where user configurations are automatically saved to lightweight configuration files, enabling rapid iteration and reproducible analysis across sessions.}\label{fig:diagram}\endgroup
\end{figure}

Computational algorithms 
 represent essential tools for extracting quantitative insights from scientific data, yet their widespread adoption remains constrained by the challenge of creating accessible interfaces. Algorithm developers often lack the time to build user-friendly graphical interfaces, while end-users may lack the programming skills to apply these methods directly. This gap keeps powerful algorithms from reaching the people who need them.

While powerful algorithms are being created, their adoption by non-programming users has been hampered by interfaces demanding programmatic expertise, creating a dichotomy between algorithm development and practical application. Recent community efforts have begun addressing this challenge, with notable successes in specific domains such as ZeroCostDL4Mic and DL4MicEverywhere for deep learning in microscopy \citep{von2021democratising,DL4MicEverywhere} and napari for interactive image visualisation \citep{sofroniew2025napari}. However, these solutions typically target specific computational environments or application domains. What remains missing is a framework enabling developers to write interface specifications once and deploy them anywhere at the same time as it automatically maintains parameter configurations that enhance reproducibility and facilitate systematic testing across different parameter combinations.

Traditional approaches to bridging this accessibility gap have typically followed two distinct paths, each with inherent limitations. The first involves developing dedicated graphical user interfaces (GUIs) for individual algorithms, exemplified by tools such as CellProfiler for bioimage analysis \citep{mcquin2018cellprofiler} and Orange for data mining \citep{demsar2013orange}. Although these provide accessible entry points for non-programmers, they require substantial additional development effort, create fragmented user experiences across tools, and often constrain users to predefined workflows that limit algorithmic flexibility. Such custom solutions also typically target a single computational environment, which prevents developers from distributing their tools to users working in different contexts. The second approach uses notebook environments such as Jupyter \citep{kluyver2016jupyter}, which offer more accessible entry points and have become the standard for exploratory data analysis. However, notebook-based approaches still necessitate direct code manipulation for parameter configuration and lack persistent parameter settings for iterative workflows. This absence of parameter memory forces users to repeatedly input configurations across sessions, diminishing productivity in exploratory research contexts where rapid iteration over parameter spaces is fundamental, and making systematic testing across parameter combinations unnecessarily cumbersome.

ImageJ and its distribution FIJI have demonstrated the value of parameter persistence through their parameter memory systems, which automatically retain user settings across sessions, enabling rapid experimental iterations that improve productivity in image analysis pipelines \citep{imagej,schindelin2012fiji}. This functionality has proven particularly valuable in microscopy workflows, where researchers frequently fine-tune processing parameters across diverse datasets. However, translating this functionality to the Python ecosystem has proven challenging due to architectural differences in development frameworks and the absence of a unified parameter management system that operates consistently across computational environments. Existing Python GUI frameworks such as ipywidgets provide building blocks for notebook interfaces but cannot be easily deployed elsewhere as they were developed with only jupyter notebooks in mind.

Here, we present EZInput, a cross-runtime environment Python library that bridges this accessibility gap by enabling algorithm developers to create user-friendly interfaces with minimal effort, making their tools accessible to end-users without programming expertise across different computational environments  . The framework employs a declarative specification system where developers define input requirements and constraints once, after which EZInput automatically manages interface rendering, validation, and parameter persistence across Jupyter notebooks, Google Colab, and terminal environments ({Fig. \ref{fig:diagram}}). This "write once, run anywhere" architecture eliminates the need for parallel development of multiple interfaces while maintaining full feature parity across environments. EZInput adapts ImageJ/FIJI's parameter memory concept to modern Python workflows through lightweight configuration files that automatically save and restore user settings across sessions, enabling the rapid iteration essential for exploratory data analysis . EZInput separates interface concerns from computational logic, so algorithm developers can focus on their scientific domain while the library handles user interaction, validation, and parameter management.

\begin{figure}[t]
\centering
\makebox[\linewidth][c]{
  \includegraphics[width=\linewidth,keepaspectratio,draft=false]{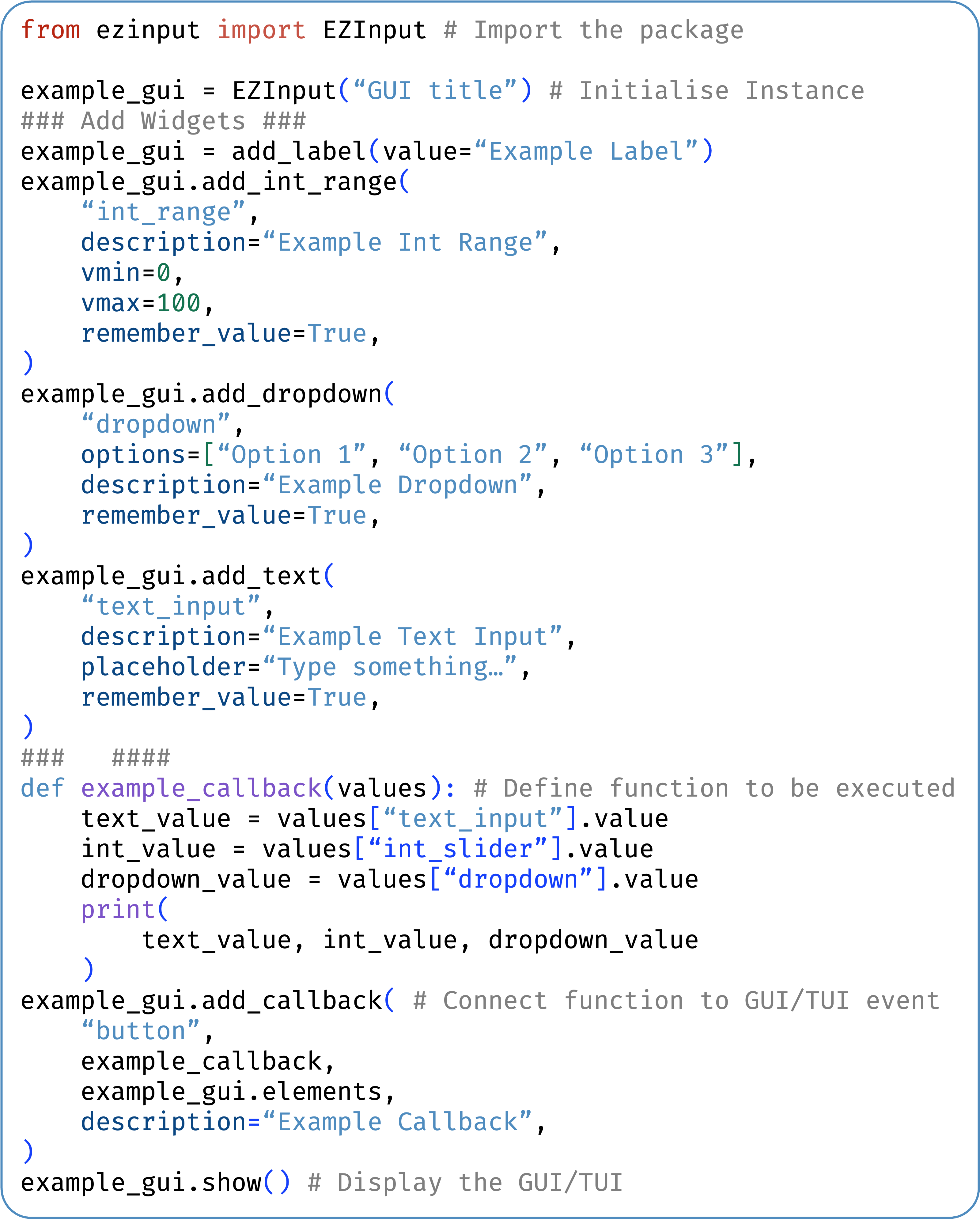}
}
\begingroup\captionsetup{width=\linewidth,singlelinecheck=false,justification=justified}\setlength{\abovecaptionskip}{6pt}\setlength{\belowcaptionskip}{6pt}\ifdefined\justifying\justifying\fi\caption{\textbf{EZInput generates cross-application user interfaces from unified declarative specifications.} The framework produces consistent interfaces across computational environments without environment-specific code. Underlying Python code demonstrates the declarative specification system that generates both interfaces. A single parameter definition block specifies labels and inputs, with EZInput automatically handling environment detection and appropriate interface rendering. This "write once, run anywhere" approach eliminates the need for parallel development of multiple interfaces while maintaining full feature parity across environments. Parameter persistence functionality ensures that user configurations remain consistent between Jupyter and terminal environments, enabling transitions between interactive exploration and production workflows.}\label{fig:cross-application}\endgroup
\end{figure}

\subsubsection{Declarative Interface Specification and Automatic Environment Detection}

EZInput implements a declarative parameter specification system that separates interface content definition from presentation logic ({Fig. \ref{fig:cross-application}}). Developers specify input requirements through a simple, intuitive API where each parameter is defined by its type (integer range, float range, text input, file path, dropdown selection, checkbox), label, constraints (minimum/maximum values, valid options, file extensions), and optional default values. The library's core architecture automatically detects the execution environment and renders appropriate interface elements without requiring environment-specific code from developers. This automatic environment detection mechanism examines the Python runtime context, checking for the presence of IPython kernel connections and interactive shell characteristics to determine the optimal rendering backend.

In Jupyter notebook environments, including Google Colab \citep{colab}, EZInput uses the ipywidgets library to generate interactive graphical controls within the notebook interface. Numerical parameters are rendered as sliders with real-time value display, dropdown selections appear as native select widgets, text inputs provide inline editing with validation feedback, and file path inputs offer interactive file browser integration  (see {Table \ref{stable:apicalls}} for complete widget type listing). The generated interfaces maintain full compatibility with Jupyter's output cell system, displaying parameter configurations inline with code cells and preserving interactive state across notebook sessions.

For terminal environments, EZInput employs prompt\_toolkit to construct text-based user interfaces (TUIs) that provide keyboard-navigable parameter configuration without graphical display requirements \citep{prompt_toolkit}. The terminal interface implements identical functionality to the Jupyter version, including parameter validation, constraint enforcement, help text display, and configuration persistence, ensuring consistent user experience regardless of computational context. Parameters are presented in a vertically scrollable list with clear visual hierarchy, keyboard shortcuts enable rapid navigation between fields, and validation feedback appears inline as users modify values. This terminal compatibility proves particularly valuable in high-performance computing environments and remote server contexts where graphical displays are limited  but user-friendly parameter configuration remains essential.

\begin{figure}[!t]
\centering
\makebox[\linewidth][c]{
  \includegraphics[width=\linewidth,keepaspectratio,draft=false]{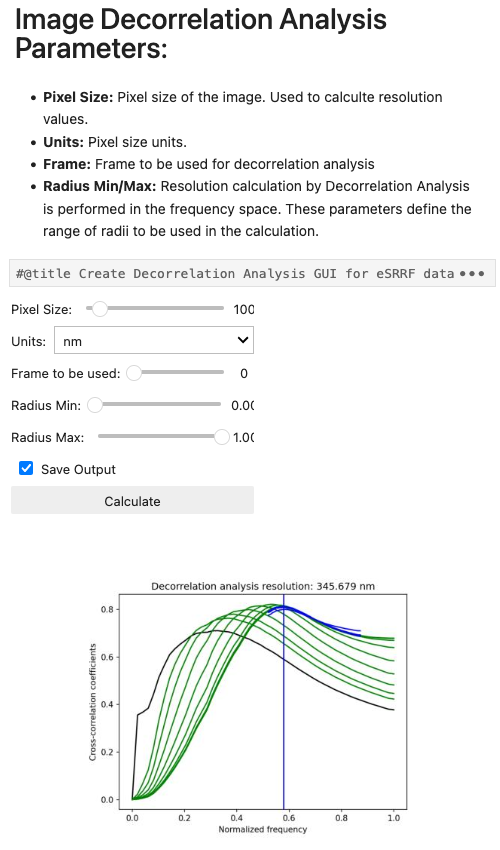}
}
\begingroup\captionsetup{width=\linewidth,singlelinecheck=false,justification=justified}\setlength{\abovecaptionskip}{6pt}\setlength{\belowcaptionskip}{6pt}\ifdefined\justifying\justifying\fi\caption{\textbf{Integration of EZInput within NanoPyx enables accessible, reproducible microscopy image analysis.} In NanoPyx, EZInput's declarative parameter specification automatically produces Jupyter notebook interfaces for complex image processing routines (image registration, denoising, super-resolution reconstruction and image quality assessment). Parameters (numeric ranges, paths, algorithm modes) persist across sessions via lightweight YAML memory, accelerating iterative tuning and ensuring identical settings can be reapplied or shared. This integration lowers the barrier for non-programming users while preserving full algorithmic flexibility, supporting the transition from interactive exploration to scripted or HPC execution using the same saved configurations.}\label{fig:nanopyx-example}\endgroup
\end{figure}

\subsubsection{Parameter Persistence and Workflow Reproducibility}

EZInput's parameter persistence system saves user configurations to lightweight YAML files stored alongside Python scripts or in user-specified locations. This functionality adapts ImageJ/FIJI's parameter memory concept to modern Python workflows, addressing a critical gap in existing notebook-based scientific computing tools. When users configure parameters through either Jupyter or terminal interfaces, EZInput automatically serializes their selections to a human-readable YAML file using a naming convention derived from the interface title. Upon subsequent executions, the library automatically detects and loads these configuration files, pre-populating interface elements with previously saved values. This eliminates the redundant parameter input that typically characterizes exploratory research workflows, where scientists iteratively refine analysis parameters across multiple sessions.

The parameter persistence mechanism implements selective memory through a flag   on individual parameters, allowing developers to specify which parameters should persist across sessions and which should reset to defaults. This granular control proves essential for workflows where certain parameters (such as file paths or experimental conditions) vary between runs while others (such as algorithmic hyperparameters or processing settings) typically remain constant. Configuration files employ a structured YAML format that maintains parameter names, values, and metadata, staying valid across parameter additions or removals between code versions. The human-readable format also enables manual editing of configuration files when programmatic parameter adjustment is desired, and supports version control integration for collaborative research contexts where parameter configurations constitute part of the experimental methods.

Beyond individual workflow efficiency, parameter persistence improves reproducibility in computational research. Configuration files generated by EZInput can be shared alongside datasets and analysis scripts, enabling collaborators to reproduce analyses with identical parameter settings without requiring detailed written documentation of each parameter value. This addresses reproducibility challenges documented in computational science \citep{sandve2013ten}, where parameter configurations often remain inadequately documented in traditional methods sections. In collaborative projects, shared configuration files ensure consistency across multiple users executing the same analysis pipeline, reducing variability introduced by manual parameter transcription errors. The lightweight nature of YAML configuration files also facilitates inclusion in supplementary materials for scientific publications, providing complete transparency of computational methods.

\subsubsection{Ecosystem Position and Distinctive Features}

EZInput addresses the accessibility barrier in computational science by enabling algorithm developers to create user-friendly interfaces with minimal development effort. Through automatic GUI generation, parameter persistence, and cross-environment consistency, the library helps developers make computational methods accessible to end-users, such as experimental biologists, who may lack programming expertise, while supporting systematic testing across environments. By requiring only declarative specification of input requirements, EZInput eliminates interface development overhead while maintaining full algorithmic flexibility, so developers can focus on their computational methods instead of GUI implementation. The "write once, run anywhere" architecture also streamlines developer workflows: the same parameter configurations run from local Jupyter notebooks to cloud platforms to HPC clusters without code modification.

The framework's success is exemplified by its integration into NanoPyx, a high-performance bioimage analysis library, where microscopy researchers use EZInput to process their images with the methods implemented in NanoPyx ({Fig. \ref{fig:nanopyx-example}}) \citep{saraiva2025nanopyx}. We have also adapted the ColabFold notebook \citep{colabfold} to use EZInput for parameter configuration, demonstrating its versatility across diverse scientific domains.

Unlike full platforms such as CellProfiler \citep{mcquin2018cellprofiler} or ImageJ/FIJI \citep{schindelin2012fiji}, EZInput is a lightweight library that integrates into existing Python codebases . Compared to GUI frameworks including ipywidgets, EZInput distinguishes itself through automatic interface generation from declarative specifications, built-in parameter persistence addressing reproducibility, and native terminal support enabling HPC deployment. While web-based frameworks excel at creating dashboards, they require server infrastructure incompatible with many HPC environments where scientific computing occurs. EZInput's minimal dependencies (ipywidgets, prompt\_toolkit, PyYAML) facilitate integration without conflicts while enabling developers to validate their tools across multiple execution contexts with identical parameter sets.

 Automatic configuration files document exact algorithmic settings, enabling inclusion in supplementary materials, sharing among collaborators for analytical consistency, and version control alongside analysis scripts. These same configuration files also enable systematic testing: developers define parameter sets once and validate behaviour across different computational environments. This addresses reproducibility concerns raised by Sandve et al. \citep{sandve2013ten} regarding inadequate documentation of computational methods as it provides a practical infrastructure for systematic validation workflows.

EZInput's declarative system suits applications with well-defined parameters that map to standard input types. Applications that need custom visualisations or real-time graphical feedback are better served by specialised frameworks with lower-level control, making EZInput complementary to visualisation tools like napari \citep{sofroniew2025napari}.

The open-source nature under MIT license facilitates community development, with ongoing priorities including enhanced documentation, cross-environment testing, and integration patterns for common scientific Python frameworks (scikit-learn, scikit-image, PyTorch).
By lowering the effort of adding a usable interface, EZInput lets developers treat accessibility as a default step instead of an afterthought .

\vspace{1em}

\begin{manuscriptinfo}
This manuscript was prepared using {\color{red}R}$\chi$iv-Maker v1.24.1~\cite{saraiva_2025_rxivmaker}. This work is licensed under CC BY 4.0.
\end{manuscriptinfo}

\begin{code}
The EZInput library is available at \url{https://github.com/HenriquesLab/EZInput}. All source code is under an MIT License. The ColabFold notebook adapted to use EZInput is available at \url{https://colab.research.google.com/github/IvanHCenalmor/ColabFold/blob/main/AlphaFold2.ipynb}. The NanoPyx library integrating EZInput is available at \url{https://github.com/HenriquesLab/NanoPyx}.
\end{code}

\begin{acknowledgements}
B.S. and R.H. acknowledge support from the European Research Council (ERC) under the European Union's Horizon 2020 research and innovation programme (grant agreement No. 101001332) (to R.H.) and funding from the European Union through the Horizon Europe program (AI4LIFE project with grant agreement 101057970-AI4LIFE and RT-SuperES project with grant agreement 101099654-RTSuperES to R.H.). Funded by the European Union. However, the views and opinions expressed are those of the authors only and do not necessarily reflect those of the European Union. Neither the European Union nor the granting authority can be held responsible for them. This work was also supported by a European Molecular Biology Organization (EMBO) installation grant (EMBO-2020-IG-4734 to R.H.), a Chan Zuckerberg Initiative Visual Proteomics Grant (vpi-0000000044 with \url{https://doi.org/10.37921/743590vtudfp} to R.H.), and a Chan Zuckerberg Initiative Essential Open Source Software for Science (EOSS6-0000000260). This study was funded by the Research Council of Finland (338537, 371287 and 374180 to G.J.), the Sigrid Juselius Foundation (to G.J.), the Cancer Society of Finland (Syöpäjärjestöt; to G.J.), and the Solutions for Health strategic funding for Åbo Akademi University (to G.J.). This research also received support from the InFLAMES Flagships Programme of the Research Council of Finland (decision numbers: 337530, 337531, 357910, and 35791). I.H. was supported by the Ministry of Education and Culture's Doctoral Education Pilot under Decision No. VN/3137/2024-OKM-6 (The Finnish Doctoral Program Network in Artificial Intelligence, AI-DOC).
\end{acknowledgements}

\begin{exauthor}
\begin{extendedauthorlist}
\extendedauthor{Bruno M. Saraiva}{\orcidicon{0000-0002-9151-5477}; \xicon{Bruno\_MSaraiva}; \linkedinicon{bruno-saraiva}}
\extendedauthor{Iván Hidalgo-Cenalmor}{\orcidicon{0009-0000-8923-568X}}
\extendedauthor{António D. Brito}{\orcidicon{0009-0001-1769-2627}}
\extendedauthor{Damián Martínez}{\orcidicon{0000-0002-5906-598X}}
\extendedauthor{Tayla Shakespeare}{\orcidicon{0000-0002-4159-0460}}
\extendedauthor{Guillaume Jacquemet}{\orcidicon{0000-0002-9286-920X}; \twittericon{guijacquemet}; \blueskyicon{guijacquemet.bsky.social}}
\extendedauthor{Ricardo Henriques}{\orcidicon{0000-0002-2043-5234}; \xicon{HenriquesLab}; \blueskyicon{henriqueslab.bsky.social}; \linkedinicon{ricardo-henriques}}
\end{extendedauthorlist}
\end{exauthor}

\section*{Bibliography}
\bibliography{03_REFERENCES}

\section*{Methods}

\subsubsection{Core Design Philosophy and Architecture}

EZInput is built on a declarative design that separates content specification from presentation logic. The library lets developers declare parameter requirements through a concise API instead of writing procedural code that constructs interface elements by hand. Developers specify the parameter type (integer range, float range, text input, file path, dropdown selection, checkbox, text area), descriptive label, validation constraints (minimum/maximum values, valid options, file extensions), and optional default values. The library then assumes responsibility for all aspects of interface rendering, including environment detection, widget creation, layout management, event handling, and state persistence. This architectural separation of concerns enables algorithm developers to focus on their scientific domain without engaging with GUI implementation complexities, thereby promoting creation of accessible computational tools.

The core architecture implements three distinct layers: the specification layer where developers define parameters, the abstraction layer that translates specifications into environment-agnostic representations, and the rendering layer that generates appropriate interface elements for the detected computational context. The specification layer uses a fluent API pattern where developers chain method calls to build complete parameter configurations. The abstraction layer maintains an internal parameter registry storing metadata including parameter names, types, constraints, current values, and persistence settings. This registry serves as the single source of truth for parameter state, ensuring consistency between interface elements and underlying data structures. The rendering layer implements separate backends for Jupyter notebook (based on ipywidgets) and terminal (based on prompt\_toolkit) environments, with a factory pattern selecting the appropriate backend at runtime based on environment detection results.

\subsubsection{Environment Detection and Backend Selection}

Automatic environment detection eliminates the need for developers to specify execution context manually or maintain parallel code paths for different environments. The detection mechanism examines multiple characteristics of the Python runtime environment to reliably distinguish between Jupyter notebook contexts, interactive terminal sessions, and non-interactive script execution. The primary detection strategy checks for the presence of an active IPython kernel by attempting to import \texttt{IPython.get\_ipython()} and inspecting the returned shell type. If a ZMQInteractiveShell is detected, indicating a Jupyter notebook or JupyterLab environment, the Jupyter backend is selected. For Google Colab environments, EZInput automatically detects the Colab runtime by attempting to import the \texttt{google.colab} module and, when detected, enables the custom widget manager through \texttt{output.enable\_custom\_widget\_manager()} to ensure full ipywidgets compatibility. For other execution contexts, including standard Python interpreters, IPython terminal sessions, or non-interactive scripts, the terminal backend is selected.

The Jupyter backend uses ipywidgets  to generate interactive controls within Jupyter's display system. Each parameter type maps to appropriate widget classes: integer and float ranges to \texttt{IntSlider} and \texttt{FloatSlider} widgets respectively, dropdown selections to \texttt{Dropdown} widgets, text inputs to \texttt{Text} or \texttt{Textarea} widgets, file paths to \texttt{Text} widgets with optional file dialog integration, and checkboxes to \texttt{Checkbox} widgets. Widgets are assembled into vertical layouts using \texttt{VBox} containers, with automatic width and styling applied for visual consistency. Widget styling uses Jupyter's built-in themes to ensure compatibility with user-selected colour schemes. Event handlers are attached to widget \texttt{observe} methods to detect value changes, trigger validation logic, and update the internal parameter registry.

The terminal backend employs prompt\_toolkit (version >=3.0.0) to construct text-based user interfaces providing keyboard-navigable parameter configuration. The implementation uses prompt\_toolkit's \texttt{Application} class to create a full-screen terminal application with custom key bindings and layout management. Parameters are rendered using prompt\_toolkit's layout components, with \texttt{FormattedTextControl} elements displaying labels and current values, and input prompts configured based on parameter types. Integer and float ranges use numeric input validation with real-time bounds checking, dropdown selections employ completion-based input with arrow key navigation, text inputs support multi-line editing where appropriate, and file paths use prompt\_toolkit's built-in path completion functionality. The terminal interface implements keyboard shortcuts for rapid navigation (Tab/Shift+Tab between fields, Enter to submit, Ctrl+C to cancel) and displays validation errors inline as users modify values.

\subsubsection{Parameter Persistence Implementation}

Parameter persistence functionality stores user configurations as YAML files in the filesystem, enabling automatic parameter restoration across sessions. The persistence system implements three core operations: configuration saving, configuration loading, and selective parameter memory. When users submit a completed parameter configuration, the library serializes all parameters marked with \texttt{remember\_value=True} to a YAML file using the PyYAML library. The default storage location places configuration files in the same directory as the executing Python script, using a naming convention derived from the interface title (spaces replaced with underscores, appended with \texttt{\_parameters.yml}). Developers can override the default location by specifying a \texttt{params\_file} path during EZInput initialization.

The YAML file structure employs a flat dictionary mapping parameter names to values, with additional metadata including timestamp of last save and parameter types for validation during loading. Example structure:

{\footnotesize
\begin{verbatim}
esrrf_order: 1
frames_per_timepoint: 250
magnification: 2
mpcorrection: true
ring_radius: 1.5
save: true
sensitivity: 1
\end{verbatim}
}

Configuration loading occurs automatically when EZInput is initialized with a specified \texttt{params\_file} or when a default configuration file is detected. The library reads the YAML file using \texttt{yaml.safe\_load()}, validates parameter names against the current parameter specification, and pre-populates interface elements with saved values. Type checking ensures loaded values match expected parameter types, with automatic conversion attempted for compatible types (e.g., integer to float) and defaults applied for incompatible or missing values. This loading mechanism keeps configuration files valid even when parameter specifications change across code versions.

Selective parameter memory through the \texttt{remember\_value} flag enables fine-grained control over persistence behaviour. Parameters with \texttt{remember\_value=True} are included in saved configurations and automatically restored. Parameters with \texttt{remember\_value=False} revert to defaults. This distinction is essential for workflows where certain parameters (such as run-specific file paths) should not persist while others (such as algorithmic hyperparameters) remain constant. The implementation stores the \texttt{remember\_value} setting in the parameter registry, consulting it during both save and load operations.

\subsubsection{Implementation Details and Dependencies}

EZInput is implemented as a pure Python package with minimal dependencies to maximise compatibility and simplify installation. Core dependencies include ipywidgets (>=8.0.0) for Jupyter interface rendering, prompt\_toolkit (>=3.0.0) for terminal interface construction, and PyYAML (>=5.0) for configuration file handling with both JSON and YAML format support. The package requires Python >=3.9 and uses modern language features including type hints, dataclasses, and match statements where appropriate.

The codebase adheres to PEP 8 style guidelines with automated formatting via Ruff, ensuring consistent code style across contributions. Type hints throughout the codebase enable static type checking via mypy, catching type-related errors during development instead of at runtime. Documentation is generated automatically from docstrings using pdoc, with all public APIs documented following Google-style docstring conventions. The project structure separates core functionality (\texttt{ezinput.py}), Jupyter-specific rendering (\texttt{ezinput\_jupyter.py}), terminal-specific rendering (\texttt{ezinput\_prompt.py}), and shared utilities, facilitating maintenance and extension. 

The library is distributed via the Python Package Index (PyPI) under the MIT license, enabling installation via \texttt{pip install ezinput}. Source code is hosted on GitHub (\url{https://github.com/HenriquesLab/EZInput}) with continuous integration workflows automating testing, documentation building, and package deployment. Pre-commit hooks enforce code quality standards including formatting, linting, and type checking before commits are accepted. Semantic versioning conventions guide release numbering, with clear changelog documentation of additions, changes, and deprecations across versions.

\onecolumn
\newpage



\renewcommand{\figurename}{Sup. Fig.}
\renewcommand{\tablename}{Sup. Table}
\setcounter{sfigure}{0}
\setcounter{stable}{0}

\newpage
\thispagestyle{empty}
\begin{center}

\vspace*{3cm}

\textbf{\Large Supplementary Information}

\vspace{3cm}

{\Huge\textbf{EZInput: A Cross-Environment Python Library for Easy UI Generation in Scientific Computing}}

\vspace{\fill}

\begin{minipage}{\textwidth}
\centering
\end{minipage}

\end{center}
\newpage

\begin{stable}[ht]
\centering
\begin{tabular}{|l|l|l|l|}
\hline
Method & Description & Jupyter & Terminal \\
\hline
\textbf{Label and Display Elements} &  &  &  \\
\hline
\texttt{add\_label} & Add non-interactive text label & yes & yes \\
\hline
\texttt{add\_HTML} & Add HTML formatted text & yes & no \\
\hline
\textbf{Text Input Methods} &  &  &  \\
\hline
\texttt{add\_text} & Single-line text input & yes & yes \\
\hline
\texttt{add\_text\_area} & Multi-line text input & yes & yes \\
\hline
\textbf{Numerical Input Methods} &  &  &  \\
\hline
\texttt{add\_int\_range} & Integer range slider/input & yes & yes \\
\hline
\texttt{add\_float\_range} & Float range slider/input & yes & yes \\
\hline
\texttt{add\_int\_text} & Integer text input with validation & yes & yes \\
\hline
\texttt{add\_bounded\_int\_text} & Bounded integer text input & yes & yes \\
\hline
\texttt{add\_float\_text} & Float text input with validation & yes & yes \\
\hline
\texttt{add\_bounded\_float\_text} & Bounded float text input & yes & yes \\
\hline
\textbf{Categorical Input Methods} &  &  &  \\
\hline
\texttt{add\_check} & Boolean checkbox/yes-no input & yes & yes \\
\hline
\texttt{add\_dropdown} & Single-selection dropdown menu & yes & yes \\
\hline
\texttt{add\_select\_multiple} & Multiple-selection widget & yes & no \\
\hline
\textbf{File System Methods} &  &  &  \\
\hline
\texttt{add\_path\_completer} & File/directory path input with completion & no & yes \\
\hline
\texttt{add\_file\_upload} & File upload widget & yes & no \\
\hline
\textbf{Interactive Elements} &  &  &  \\
\hline
\texttt{add\_callback} & Callback function with button trigger & yes & yes \\
\hline
\texttt{add\_output} & Output display area & yes & no \\
\hline
\textbf{Custom Widgets} &  &  &  \\
\hline
\texttt{add\_custom\_widget} & Add custom ipywidgets widget & yes & no \\
\hline
\textbf{Configuration Management} &  &  &  \\
\hline
\texttt{save\_parameters} & Save parameters to specified file & yes & yes \\
\hline
\texttt{load\_parameters} & Load parameters from specified file & yes & yes \\
\hline
\texttt{get\_values} & Retrieve current parameter values & yes & yes \\
\hline
\end{tabular}
\raggedright
\caption{\textbf{Methods implemented in EZInput and their availability across interface types.} yes indicates full support, no indicates not available on that runtime environment.}
\label{stable:apicalls}
\end{stable}

\end{document}